\documentclass{PoS}

\usepackage{amsmath,amsfonts}
\usepackage{import}
\usepackage{listings}

\usepackage{url}
\usepackage{tabularx}
\usepackage{booktabs}
\usepackage{slashed}
\def\SymbReg{\raisebox{0.4em}{\scalebox{0.5}{\textsuperscript{\bf\textregistered}}}{ }}
\def\SymbTra{\raisebox{0.3em}{\scalebox{0.5}{\textsuperscript{\bf\texttrademark}}}{ }}

\title{HISQ inverter on Intel\SymbReg Xeon Phi\SymbTra and NVIDIA\SymbReg GPUs}

\ShortTitle{HISQ inverter on Intel\SymbReg Xeon Phi\SymbTra and NVIDIA\SymbReg GPUs}

\author{O. Kaczmarek, C. Schmidt and P. Steinbrecher\\
        Fakult\"at f\"ur Physik, Universit\"at Bielefeld, D-33615 Bielefeld, Germany\\
        E-mail: \email{okacz, schmidt, p.steinbrecher@physik.uni-bielefeld.de}}

\author{\speaker{Swagato Mukherjee}\\
        Physics Department, Brookhaven National Laboratory, Upton, NY 11973, USA\\
        E-mail: \email{swagato@bnl.gov}}

\author{M. Wagner\\
        Physics Department, Indiana University, Bloomington, IN 47405, USA\\
        E-mail: \email{mathwagn@indiana.edu}}

\abstract{The runtime of a Lattice QCD simulation is dominated by a small kernel, which calculates the product of a vector by a sparse matrix known as the ``Dslash'' operator. Therefore, this kernel is frequently optimized for various  HPC architectures. In this contribution we compare the performance of the Intel\SymbReg Xeon Phi\SymbTra to current Kepler-based NVIDIA\SymbReg Tesla\SymbTra GPUs running a conjugate gradient solver. By exposing more parallelism to the accelerator through inverting multiple vectors at the same time we obtain a performance $>\!\!250\;\mathrm{GFlop/s}$ on both architectures. This more than doubles the performance of the inversions. We give a short overview of both architectures, discuss some details of the implementation and the effort required to obtain the achieved performance.}

\FullConference{The 32nd International Symposium on Lattice Field Theory\\
                23-28 June, 2014\\
                Columbia University New York, NY}
\begin{document}

\section{Introduction}

For the analysis of QCD simulations one often needs to perform many inversions of the Fermion Matrix for a constant gauge field. In finite temperature QCD the calculation of the fluctuation of conserved charges, brayon number (B), electric charge (Q) and strangeness (S) is such an example. Their calculation is particular interesting as they can measured in experiments at the Relativistic Heavy Ion Collider (RHIC) and the Large Hadron Collider (LHC) and also be determined from generalized susceptibilities in Lattice QCD
\begin{align}
    \chi^{BQS}_{mnk}(T) =  \left.\frac{1}{VT^3}  
    \frac {\partial^{m+n+k} \ln \mathcal{Z}} {
    \partial \left( \mu_B / T \right)^m
    \partial \left( \mu_Q / T \right)^n
    \partial \left( \mu_S / T \right)^k
    }
    \right|_{\vec{\mu}=0}\ ,
\end{align}
where $\mathcal{Z}$ denotes the partition function of the medium at temperature $T$ and volume $V$. The required derivatives w.r.t.\ the chemical potentials $\mu$ can be obtained by stochastically estimating the traces with a sufficiently large number of random vectors $\eta$, e.g.
\begin{align}
\operatorname{Tr} \left(\frac{\partial^{n_1} M}{\partial \mu^{n_1}} M^{-1} \frac{\partial^{n_2} M}{\partial \mu^{n_2}} \ldots M^{-1}\right) = \lim_{N \rightarrow \infty} \frac{1}{N} \sum_{k=1}^{N} \eta_k^\dagger \frac{\partial^{n_1} M}{\partial \mu^{n_1}} M^{-1} \frac{\partial^{n_2} M}{\partial \mu^{n_2}} \ldots M^{-1} \eta_k\ .
\end{align}
For each random vector we need to perform several inversions of the Fermion Matrix $M$, depending on the highest degree of derivative we calculate. Typically we use 1500 random vectors to estimate the traces on a single gauge configuration. To also reduce the gauge noise we need between 5000 and 20000 gauge configurations for each temperature. The generated data have been used for several investigations in the last years~\cite{Bazavov:2014yba_xya}.

For reasons of the numerical costs, staggered fermions are the most common type of fermions for thermodynamic calculations on the lattice. We use the highly improved staggered fermion (HISQ) action~\cite{Follana:2004HISQ}. It reduces the taste-splitting as much as possible.
The HISQ action uses two levels of fat7 smearing and a Naik term. In terms of the smeared links $X$ and Naik links $N$ the Dslash operator reads
\begin{align}
	w_x = D_{x,x'} v_{x'} 
	= \sum_{\mu=0}^4
	 \left[ \left (X_{x,\mu} v_{x+\mu}- X^\dagger_{x-\mu,\mu } v_{x-\mu}\right) + 
	 \left(N_{x,\mu} v_{x+3\mu}- N^\dagger_{x-3\mu,\mu } v_{x-3\mu}\right)
	 \right]\ .
\end{align}

In the Krylov solvers used for the inversion the application of the Dslash operator is the dominating term. It typically consumes more than $80\%$ of the runtime. It has a low arithmetic intensity (see table~\ref{arithm_intens}). Hence, the performance is bound by the available memory bandwidth. These types of problems are well suitable for accelerators as these currently offer memory bandwidths in the range of $200-400\;\mathrm{GB/s}$ and with the use of stacked DRAM are expected to reach $1\;\mathrm{TB/s}$ in the next years. Still the most important factor when tuning is to avoid memory transfers. A common optimization is to exploit available symmetries and reconstruct the gauge links from 8 or 12 floats instead of loading all 18 floats. For improved actions these symmetries are often broken. For the HISQ action only the Naik links can be reconstructed from $9$ or $13/14$ floats.

For our and many other applications a large number of inversions are performed on a single gauge configuration. Then one can exploit the constant gauge field by grouping the random vectors in small bundles, thus applying the Dslash for multiple right-hand sides (rhs) at once:
\begin{equation}
	\left(w_x^{(1)}, w_x^{(2)}, \ldots, w_x^{(n)}\right) = D_{x,x'} \left(v_{x'}^{(1)}, v_{x'}^{(2)}, \ldots, v_x^{(n)}\right)\ .
\end{equation}
%
This increases the arithmetic intensity of the HISQ Dslash as the loading of the gauge field occur only once for the $n$ rhs. 
\begin{table}
    \centering
    \begin{tabular}{cccccccccc}\toprule
        \#rhs && 1 & 2 & 3 & 4 & 5 &6 & 8\\ \midrule
        $\mathrm{Flop/byte}$ (full)&& 0.73 & 1.16 & 1.45 & 1.65 & 1.80 & 1.91 & 2.08 \\ 
        $\mathrm{Flop/byte}$ (r14)&&0.80 &1.25 & 1.53 & 1.73 & 1.87 &1.98 & 2.14\\\bottomrule
    \end{tabular}
    \caption{\label{arithm_intens}The arithmetic intensity of the HISQ Dslash for different number of right-hand sides using full or reduced 14 float storage (r14) for the Naik links.}
    \vspace*{-0.5em}
\end{table}
Throughout the following we will only present performance for single precision computations. Increasing the number of rhs from 1 to 4 already results in an improvement by a factor of more than 2. For even higher $n$ the relative effect is less significant. In the limit $n \rightarrow \infty$ the highest possible arithmetic intensity that can be reached is $\sim\! 2.75$. At $n=8$ we have reached already $\sim\! 75\%$ of the limiting peak intensity while for 1 rhs we only obtain $25\! -\! 30\%$. It is also obvious that for an increasing number of rhs the memory transfers caused by loading the gauge fields is no longer dominating and the impact of reconstructing the Naik links reduces from $\sim\! 10\%$ for a single rhs to $\sim\! 3\%$ for 8 rhs. For the full conjugate gradient the additional linear algebra does not allow for the reuse of any constant fields. The effect of the increased arithmetic intensity of the Dslash will therefore be less pronounced in the full CG.

\begin{table}[h]
    \centering
    \begin{tabular}{lcccc} \toprule
        & Phi\SymbTra 5110P & K20 & K40 & GTX\SymbTra Titan
          \\\midrule
        Cores / SMX 				& 60 	& 13 	& 15  & 14  \\
        (Threads/Core) / (Cores/SMX)	& 4 	& 192 & 192 & 192\\
        Clock Speed [$\mathrm{MHz}$]	& 1053 	& 706 	& 745/810/875 & 837 \\
        L1 Cache / Core [$\mathrm{KB}$]	& 32	& 16-48 & 16- 48 & 16 - 48 \\
        L2 Cache [$\mathrm{MB}$]		& 30	& 1.5	& 1.5 & 1.5 \\
        Memory Size [$\mathrm{GB}$]		& 8     & 5     & 12 & 6 \\
        peak fp32/64 [$\mathrm{TFlop/s}$]	& 2.02/1.01 & 3.52/1.17	& 4.29/1.43   & 4.5/1.5\\
        Memory Bandwidth [$\mathrm{GB/s}$]	& 320  & 208 & 288 & 288\\
        TDP [$\mathrm{W}$]			& 225 & 225 & 235 & 250 \\
        \bottomrule
    \end{tabular}
    \caption{Summary of the important technical data of the accelerators we have used in our benchmarks.}
\end{table}

In the following we will discuss our implementation of the CG inverter for the HISQ action on NVIDIA\SymbReg GPUs and the Intel\SymbReg Xeon Phi\SymbTra. The GPUs are based on the Kepler\SymbTra architecture. The Xeon Phi\SymbTra is based on the Knights Corner architecture.

\section{GPU}

NVIDIA's current architecture for compute GPUs is called Kepler, the corresponding chip GK110. The latest compute card, the Tesla K40, comes with a slightly modified version GK110B and GPU Boost. The latter allows the user to run the GPU at a higher core clock. As memory-bandwidth bound problems usually stay well within the thermal and power envelopes the card is capable of constantly running at this higher clock for Lattice QCD simulations. The memory clock remains constant and thus a performance impact on bandwidth-bound applications is not obvious. However, the higher core clock allows to better saturate the available bandwidth. We will only show results with the highest possible core clock for the K40.
\begin{figure}[h]
    \centering
    \includegraphics[width=0.49\textwidth]{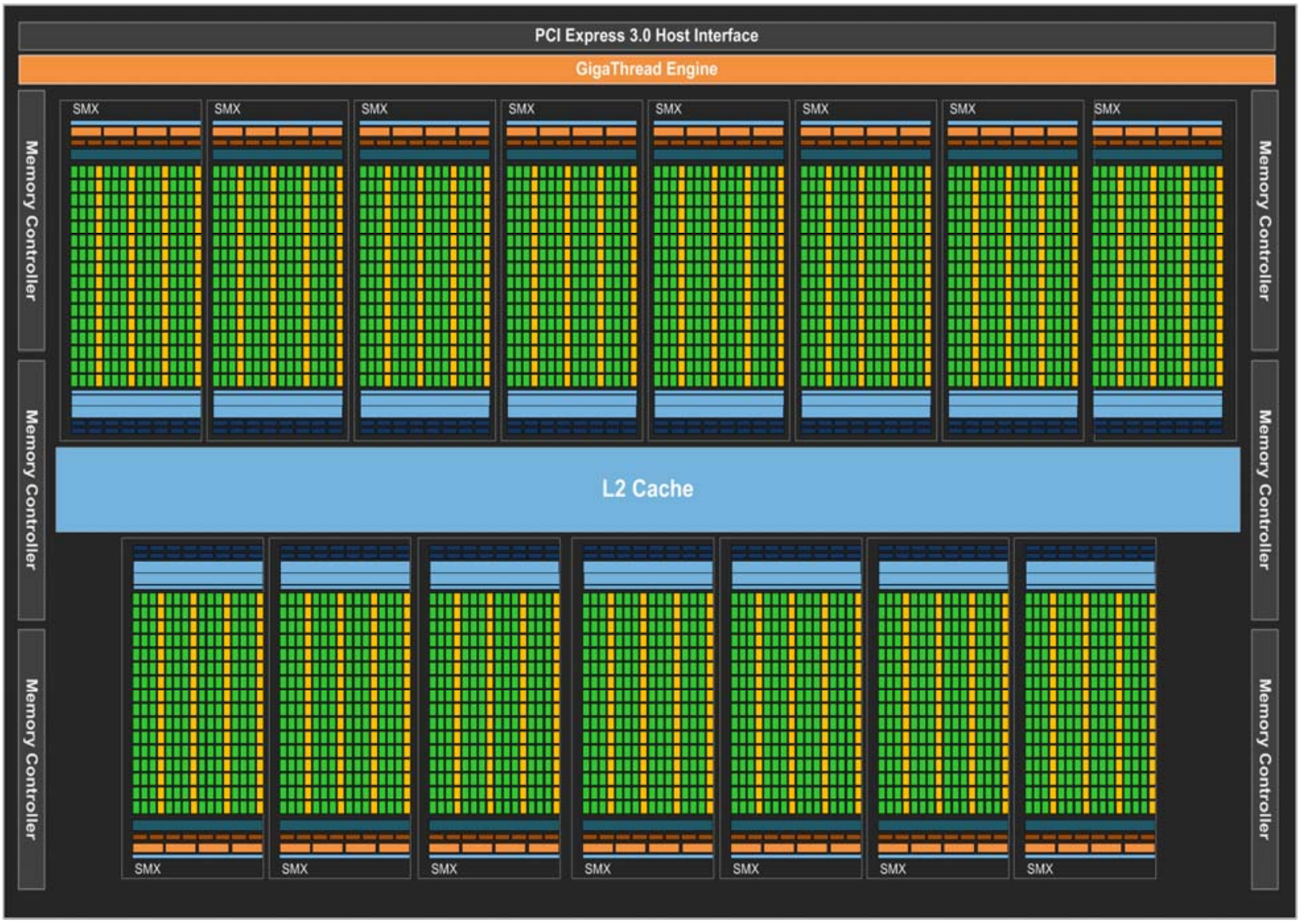}\hspace{0.01\textwidth}
    \includegraphics[width=0.3086\textwidth]{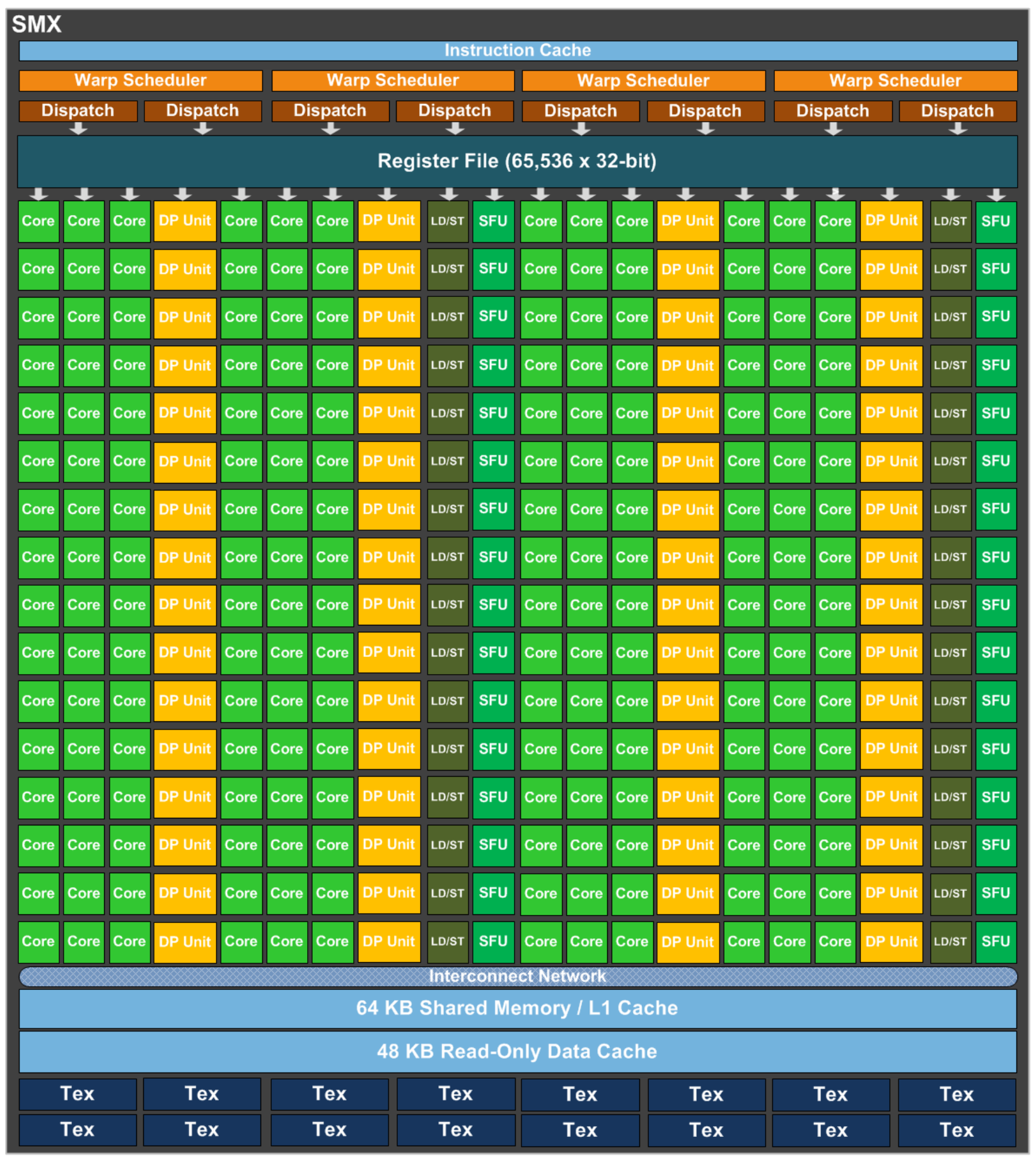}
    \caption{\label{gk110}The GK110 chip and one of its SMX processors~\cite{gk110w}.}
\end{figure}
The GK110 chip consists of several Streaming Multiprocessors (SMX). The number of SMX depends on the card. Each SMX features $192$ CUDA cores. Within a SMX they share a configurable L1 cache / shared memory area and a $48\;\mathrm{KB}$ read-only/texture cache. The shared L2 cache for all SMX is relatively small with only $1.5\;\mathrm{MB}$.

{\bf Implementation:} For a bandwidth-bound problem the memory layout and access is crucial to achieve optimal performance. As GPUs have been used for several years to accelerate Lattice QCD simulations a lot of the techniques we used may be considered as standard by now. We use a Struct of Arrays (SoA) memory layout for both the color vectors and the gauge links. We reconstruct the Naik links from 14 floats. We observed best results by loading the gauge links through the texture unit and the color vectors through the standard load path (L1).

For the implementation of the Dslash for multiple rhs two approaches are possible. On the GPU the obvious parallelization for the Dslash for one rhs is over the elements of the output vector, i.e., each thread processes one element of the output vector. The first approach (register-blocking) lets each thread multiply the already loaded gauge link to the corresponding element of multiple right-hand sides. The thread thus generates the element for one lattice site for several output vectors. This approach increases the number of registers needed per thread and will result in a lower GPU occupancy or at some point spilling. Both effects will limit the achievable performance, while the latter is less likely as each thread can use up to 255 registers for the Kepler architecture.

The second approach (texture cache blocking) is to let each thread process one element of one output vector and group the threads into two-dimensional CUDA blocks with lattice site~$x$ and rhs~$i$. As one CUDA block is executed on one SMX this ensures temporal locality for the gauge links in the texture cache. Ideally the gauge links only need to be loaded from the global memory for one rhs. When the threads for the other rhs are executed they are likely to obtain the gauge links from cache.
This approach does not increase the register pressure. Furthermore the total number of threads is increased by a factor $n$ and this may furthermore improve the overall GPU usage.

Both approaches can also be combined and the best possible solution is a question of tuning. For our benchmarks we determine the optimal configuration for a given lattice size and number of rhs for each GPU. Furthermore we employ an automatic tuning to select the optimal launch configuration for the Dslash operation also depending on the GPU, lattice size and number of rhs.

The remaining linear algebra-Kernels are kept separate for the different rhs. This allows us to easily stop the solver for individual rhs that have already met the convergence criterion. To hide latencies and allow for a better usage of the GPU we use separate CUDA streams for each right-hand side. We keep the whole solver on the GPU and only communicate the set of residuals for all rhs at the end of each iteration. 

\section{MIC}

\begin{figure}[h]
    \centering
    \includegraphics[width=0.87\textwidth]{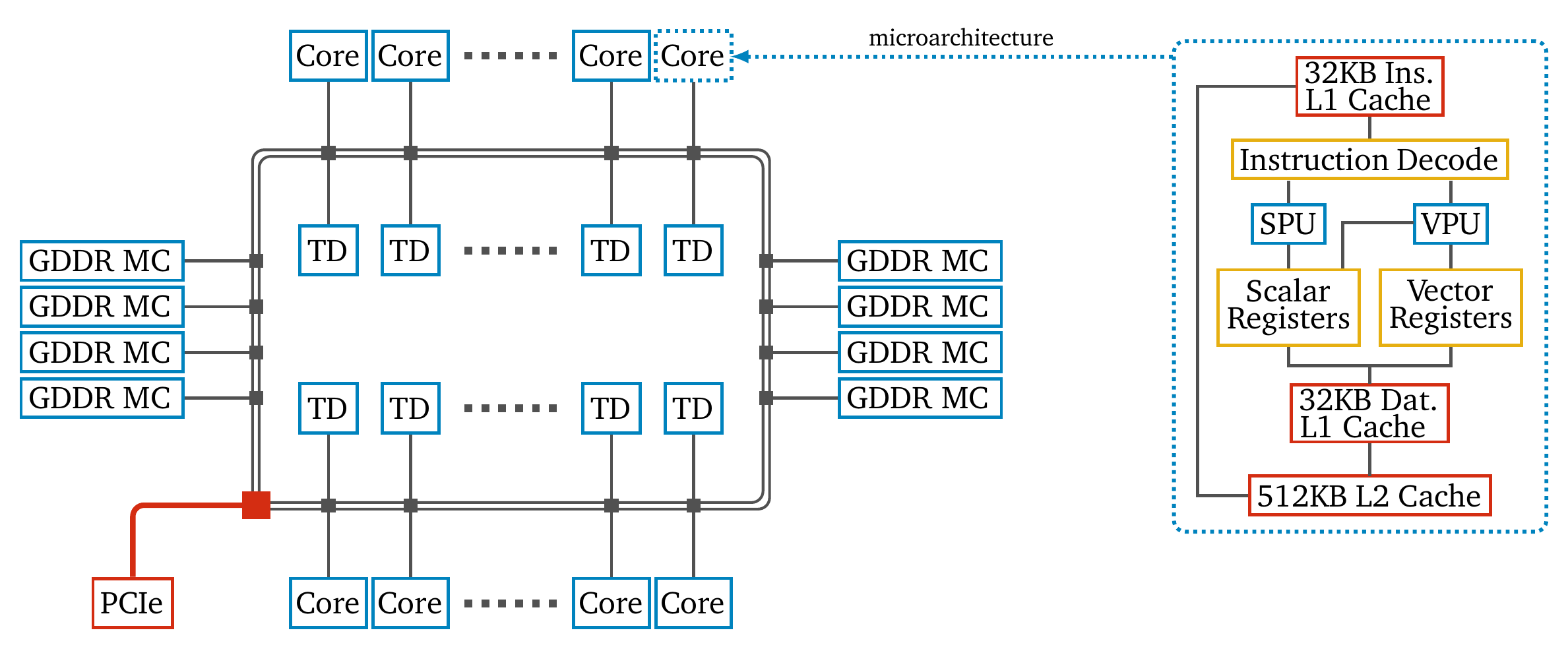}
    \caption{\label{mic_ring} Visualization of the bidirectional ring on the die (l.h.s.) and the microarchitecture of one core (r.h.s.) showing the Scalar Processing Unit (SPU), Vector Processing Unit (VPU) and the cache hierarchy. The latter is kept fully coherent through global distributed tag directories (TD).}
\end{figure}

The Intel\SymbReg Xeon Phi\SymbTra is an in-order \texttt{x}86 based many-core processor~\cite{mic}. The accelerator runs a Linux $\mu$OS and can have up to 61 cores combined via a bidirectional ring (see figure~\ref{mic_ring}). Therefore, the memory transfers are limited by concurrency reaching only $140\;\mathrm{GB/s}$ for a stream triad benchmark~\cite{stream_mic}. Each core has a private L1 data and instruction cache, as well as a global visible L2 cache. In the case of an local L2 cache miss a core can cross-snoop another's core L2 cache. If the needed data is present it is send through the ring interconnect, thus avoiding a direct memory access.
One core has thirty-two $512\;\mathrm{bit}$ \texttt{zmm} vector registers and 4 hardware context threads delivering two executed instructions per cycle while running with at least two threads per core. In order to fully utilize the Many Integrated Core (MIC) it is mostly required to run with four threads per core. Especially for memory-bound applications using four threads offers more flexibility to the processor to swap the context of a thread, which is currently stalled by a cache miss. The MIC has support for streaming data directly into memory without reading the original content of an entire cache line, thus bypassing the cache and increasing the performance of algorithms where the memory footprint is too large for the cache~\cite{mic}.

{\bf Implementation:} We have parallelized our program with OpenMP and vectorized it using low-level compiler functions  called intrinsics. Those assembly-coded functions are expanded inline and do not require explicit register management or instruction scheduling through the programmer as in pure assembly code. Also using intrinsics, the software, has to be designed in a register aware manner; only the explicit management of the registers is taken over by the compiler. We found that the compiler is only able to optimize code over small regions. Thus, the order of intrinsics has an influence on the achieved performance, thereby making optimizations more difficult. Due to the different links needed for the nearest and third-nearest neighbor term we implemented both in separate kernels, thereby reducing cache pollution and simplifying cache reuse for the vectors. 

{\bf Site fusion:} One problem of using $512\;\mathrm{bit}$ registers involving $\mathrm{SU}(3)$ matrix-vector products is that one matrix/vector does not fit into an integer number of \texttt{zmm} registers without padding. Because of that, it is more efficient to process several matrix-vector products at the same time using a site fusion method. A naive single-precision implementation could be to create a ``Struct of Arrays'' (SoA) object for 16 matrices as well as for 16 vectors. Such a SoA vector object requires 6 \texttt{zmm} registers when it is loaded from memory. One specific register then refers to the real or imaginary part of the same color component gathered from all 16 vectors, thus each vector register can be treated in a ``scalar way''. These SoA objects are stored in an array using a site ordering technique. Our Dslash kernel runs best with streaming through $xy$-planes and is specifically adapted for inverting multiple right-hand sides. Therefore, we use a 8-fold site fusion method, combining 8 sites of the same parity in $x$-direction, which makes the vector arithmetics less trivial and requires explicit in-register align/blend operations. By doing so we reduce the register pressure by 50\% compared to the naive 16-fold site fusion method.

{\bf Prefetching:} For indirect memory access, i.e.\ the array index is a non-trivial calculation or loaded from memory, it is not possible for the compiler to insert software prefetches. The MIC has a L2 hardware prefetcher which is able to recognize simple access pattern. We found that it does a good job for a linear memory access. Thus, there is no need for software prefetching by hand inside the linear algebra operations of the CG. However, the access pattern of the Dslash kernel is too complicated for the hardware prefetcher. Therefore, it is required to insert L1 and L2 software prefetches using intrinsics. The HISQ inverter runs $2\times$ faster with inserted software prefetches. 

%
%
%
%

\section{Comparison}

We performed our benchmarks for a single accelerator. We used CUDA 6.0 for the GPU and the Intel\SymbReg Compiler 14.0 for MIC. We used the default settings of MPSS 3.2, huge pages and a balanced processor affinity. First we discuss the performance as a function of the number of rhs. The maximum number of rhs is furthermore limited by the memory requirements. We observe roughly the expected scaling from the increased arithmetic intensity. When comparing the results for four right-hand sides to one right-hand side we see improvements by a factor of about 2, close to the observed increase in arithmetic intensity for the Dslash. For the full CG the linear algebra operations were expected to weaken the effect of the increased arithmetic intensity for the Dslash.
\begin{figure}[h]
    \centering
    \includegraphics[width=.497\textwidth]{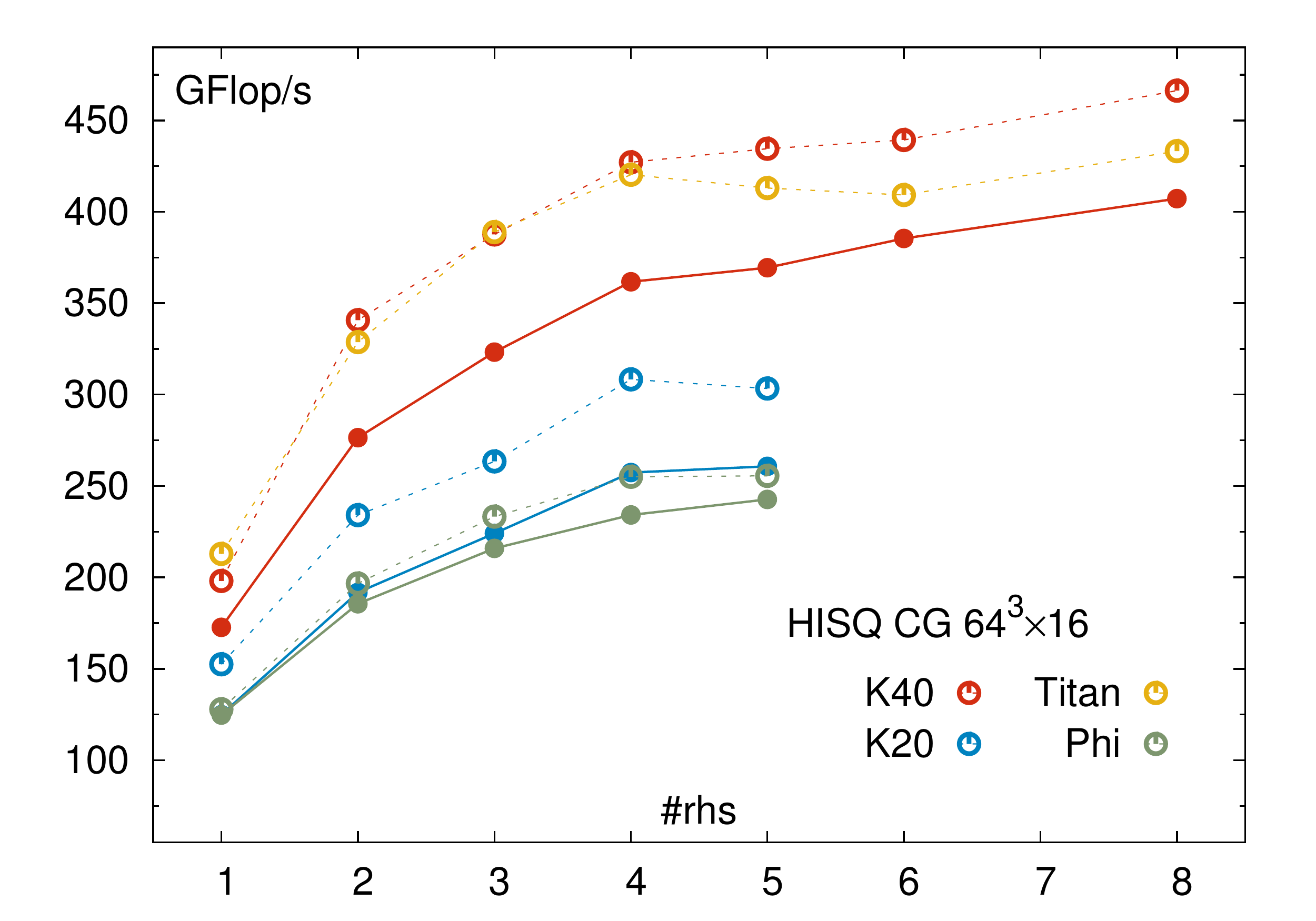}
    \includegraphics[width=.497\textwidth]{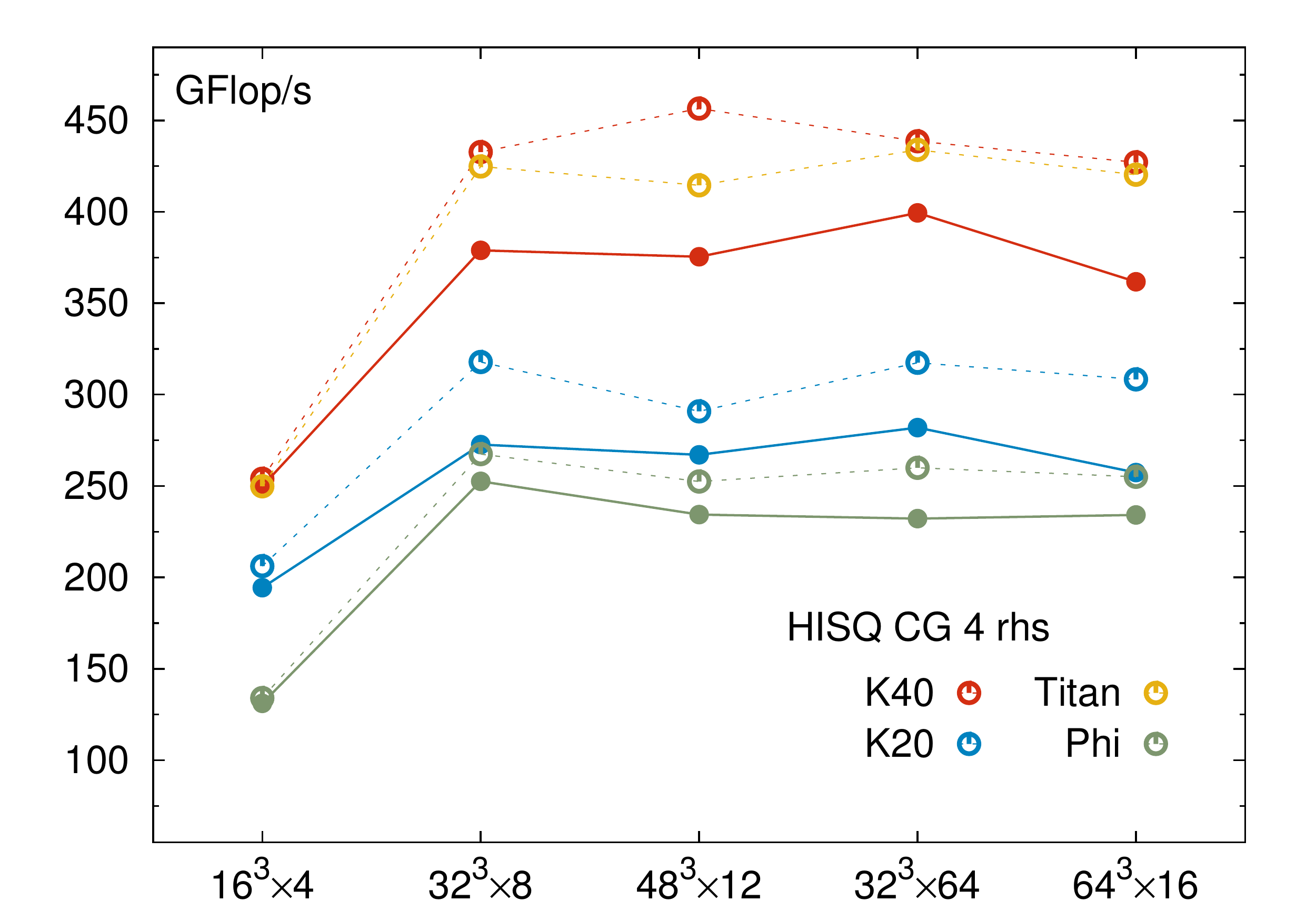}
    \caption{Performance of the HISQ inverter on different accelerators for a $64^3\!\!\times\!\!16$ lattice as a function of the number of rhs (left) and for 4 rhs as a function of the lattice size (right). We reconstruct the Naik links from 14 floats on the GPU. The dashed lines corresponds to ECC disabled devices.}
\vspace*{-1.69mm}
\end{figure}
At four right-hand sides we obtain about $10\%$ of the theoretical peak performance. For the Xeon Phi it is a bit more (${\sim\!12\%}$), but its theoretical peak performance is also the lowest of all accelerators used here while its theoretical memory bandwidth is the highest. For the GPUs we observe a performance close to the naive estimate (arithmetic intensity)~$\times$~(theoretical memory bandwidth). As previously reported the theoretical memory bandwidth is nearly impossible to reach on the Xeon Phi\SymbTra. However, our performance numbers are in agreement with the estimate (arithmetic intensity)~$\times$~(memory bandwidth from the stream benchmark). If we consider the performance of the CG for a fixed number of rhs (here $n=4$) we observe that the best performance is only obtained for lattice sizes larger than $32^3\! \times\! 8$. With our relatively new MIC code the Xeon Phi\SymbTra is slightly slower than a K20. The K40 is another $30-40\%$ faster. That is consistent with the increase in the theoretical memory bandwidth. For cases where disabled ECC is acceptable the significantly cheaper gaming~/~amateur card GTX\SymbTra Titan achieves a performance close to the K40.

{\bf Energy consumption:} A further point that we quickly checked was the typical energy consumption of the accelerator. For four right-hand sides and lattice sizes between $32^3\!\times\! 8$ and $64^3\!\!\times\!\! 16$ we observed values $\sim\!\!125\;\mathrm{W}$ for the K20 and $\sim\!\!185\;\mathrm{W}$ for the K40 without ECC. The Xeon Phi\SymbTra consumed the most energy at about $200\;\mathrm{W}$. These numbers have been measured using the system~/~accelerator counters and do not include a host system. The resulting efficiency for the Kepler architecture is hence about $2.25\;\mathrm{(GFlop/s)/W}$. For the Xeon Phi\SymbTra we estimate $\sim\!\!1.5\;\mathrm{(GFlop/s)/W}$ at four right-hand sides.

{\bf Acknowledgments:} We acknowledge support from NVIDIA\SymbReg through the CUDA Research Center program. We thank Mike Clark for providing access to a GTX\SymbTra Titan card for benchmarks. Furthermore, we would like to thank the Intel\SymbReg Developer team for their constant support. One of the authors SM is supported through the Contract No.\ DE-AC02-98CH10886 with the U.S.\ Department of Energy.

\end{document}